# Detection of Coma Activity of the ACO/Quasi-Hilda Object, 212P/2000YN$_{30}$

Y.-C. Cheng[1] and W.-H. Ip[1,2,3]
[1]Institute of Astronomy
[2] Institute of Space Science
National Central University, Taiwan
[3] Space Science Institute
Macau University of Science and Technology, Macau

Abstract

The quasi-Hilda object, 212P/2000YN$_{30}$ with a cometary-like orbit, was found to display a dust tail structure between January and March, 2009. From orbital calculations, it is shown that this object could have been an active comet in its past history before being transported to the current orbital configuration in quasi-stable 3:2 resonance with Jupiter.

I. Introduction

One basic question in connection to the origin of life and development of biosphere has to do with the source of the terrestrial ocean. For exogenic sources, the ocean water could have come from either comets or asteroids. The idea that the cometary water ice could be the main contributor (Chyba, 1987; Ip and Fernandez, 1988) has suffered a setback because the D/H isotope ratios measured in several comets are about a factor of 2 larger than the standard value $1.49\pm0.03\times10^{-4}$ of the ocean water (Eberhardt et al., 1995; Meier et al., 1998; Crovisier et al., 2004; Jehin et al., 2009). But the recent report of the D/H ratio (= $1.61\pm0.24\times10^{-4}$) of the Jupiter family comet 103P/Hartley 2 determined from the Herschel observations has brought new light to the old issue of the terrestrial water (Hartogh et al., 2011). The most recent report of D/H ratio of $2.06\pm0.22\times10^{-4}$ for the Oort cloud comet C/2009 P1 (Garradd) by

Bockelee-Morvan et al. (2012) has further underscored the point that comets could have a wider range of D/H ratios than previously thought. In any event, they could still make contribution to the terrestrial water source reservoir within the dynamical constraint of being no more than 10% in total as estimated by Morbidelli et al. (2000). We note that among the other seven comets which D/H ratios having been measured by pre-Herschel spacecraft observations or ground-based observations, C/1995 O1 Hale-Bopp, C/1996 B2 Hyakutake, C/2001 Q4 NEAT and C/2002 T7 LINEAR are all long period comets from the Oort cloud, and 1P/Halley, 8P/Tuttle and 81P/Wild2 are short-period comets (see Jehin et al., 2009). The D/H ratios of the long-period comets - which presumably have a different condensation region from that of the Jupiter-family comets - now range from $2.06 \times 10^{-4}$ (Bockelee-Morvan et al., 2012), $3.3 \times 10 \pm 0.8 \times 10^{-4}$ (Meier et al., 1998) to $4.6 \pm 1.4 \times 10^{-4}$ for C/2001 Q4 NEAT (Weaver et al., 2008). These values overlap the range between $1.49 \pm 0.03 \times 10^{-4}$ for 103P/Hartley 2 (Hartogh et al., 2011) and $4.09 \pm 10^{-4}$ for 8P.Tuttle (Villanueva et al., 2009)

Another basic question of equal importance is, therefore, how would Jupiter family comets (JFCs) presumably originated from the transneptunian region (Duncan et al., 1988; Ip and Fernandez, 1991; Volk and Malhotra, 2008) have such large difference in the D/H ratios? Could they have formed in different orbital positions in the outer solar nebula before being swept up by planet migration (Fernandez and Ip, 1984; Malhotra, 1995), Tiscareno and Malhotra (2003), Di Sisto and Brunini (2007) and Volk and Malhotra (2008) considered the dynamical transformation of the transneptunian objects (TNOs) into the Centaur populations with orbits in the Jovian and Saturnian zone and then the short-period comets in the inner solar system. Or, some of them could have formed in the asteroidal region but were later implanted in the Kuiper belt after following a complex history of dynamical evolution. Di Sisto et al. (2005) studied the interesting issue on whether JFCs could be mixed with the Hilda asteroids and vice versa by performing numerical orbital calculations on stably trapped population in 3:2 resonance with Jupiter and those quasi-stable population. They found that the majority (~99%) of the escaping Hildas would be perturbed into JFC-like orbits but with perihelia q > 2.5 AU. This means that only a small fraction (~1%) of them would become near-Earth objects with the possibility of hitting the Earth according to this study. Whether the Hilda population could have significantly contributed to the terrestrial ocean water thus depends very much on the total mass of the original population. Di Sisto et al. (2005) further estimated that at the present time the expected number of comet-sized Hilda asteroids in cometary-like orbits should be on the order of 143 in comparison to about 2800 dormant JFCs.

In a series of papers, Dahlgren and Lagerkvist (1995) and Dahlgren et al. (1997, 1999) reported on their photometric survey of the color distribution of the Hilda asteroids. In one study, they showed that about 36% of the Hildas could be classified as D-type, 28% as P-type and only 2% as C-type (Dahlgren et al., 1999). That both JFCs and Hilda asteroids share many of the characteristics of the D-type taxonomic class is another possible evidence for a link between these two populations of small bodies. From this point of view, we might pose the hypothesis that comet 103P/Hartley 2 could be an escapee from the Hilda group in case the D/H ratio of the Hildas is found to be the same as those of the CI-CM chondrites (Robert, 2006).

The above discussion shows the new impetus in the study of asteroids in cometary orbits (or ACOs) since they can provide hints on the nature and relationship between the terrestrial water and plantesimals formed in the early solar system. This is also why the recent finding of a small population of main-belt comets or active asteroids (Hsieh and Jewitt, 2006; Jewitt, 2012) has drawn so much attention. The main belt objects discovered by these authors are all located at the outer edge of the main asteroidal belt with semi-major axes between 3.156 AU and 3.196 AU and near-zero eccentricity. The detection of water ice and organics on the surface of the asteroid 24 Themis has further heightened the interest on this issue (Campins et al., 2010; Rivkin and Emery, 2010).

In our observational program at Lulin, the Tisserand parameter,

$$T_J = a_J/a + 2 \cdot \sqrt{(a/a_J)(1-e^2)} \cos i$$

has been used to select the targets. In the above equation, $a_J$ is the semi-major axis of Jupiter, and e and i are the eccentricity and inclination, respectively, of the object in question. Vaghi (1973) examined the $T_J$ values of the 73 JFCs then known and found that the majority of them fall in the range of $2.450 < T_J < 3.032$, and further concluded that those with $T_J > 2\sqrt{2}$ must have originated from elliptical orbits instead of parabolic orbits. Kresak (1980) discussed variation of the values of the Tisserand parameter among different groups of solar system small bodies and showed that the Hilda asteroids shared the same dynamical characteristics as that of the JFCs even though not a single asteroidal-like object could be definitely identified as an ex-comet at that time. Recent study by Fernandez et al. (2005) comparing the albedos (α) and $T_J$ values of 26 asteroids in cometary orbits (including 6 Damocloids and 6 near-Earth asteroids) found the interesting result that there is a discontinuity in the albedo distribution as a function of $T_J$. That is, those with $T_J < 3.0$ would tend to have $\alpha \sim 0.04$ and those with $T_J > 3.0$ would have $\alpha > 0.2$ thus suggesting different physical origins. For these reasons, we are mainly interested in objects with $T_J < 3$.

In addition to the ACOs and Hilda asteroids, many objects called quasi-Hildas moving in orbital region of the Hildas but without being stably trapped in the 3:2 mean motion resonance with Jupiter are also included in the Lulin target list according to their $T_J$ values. While Di Sisto et al. (2005) suggested that the Hildas and quasi-Hildas could be closely related to JFCs according to their dynamical behaviour, Toth (2006) gave an update on the orbital properties of ecliptic comets in Hilda-like orbits and quasi-Hilda asteroids. Subsequently Ohtsuka et al. (2009) showed that some of the quasi-Hildas (i.e., 147P/Kushida-Muramatsu) could be the progenitors of temporarily captured irregular satellites of Jupiter. Along the same line, Fernandez and Gallardo (2002) used orbital integration method to trace the dynamical evolution of inactive JFCs and estimated that up to 20% of the near-Earth asteroid population with $T_J < 3$ could be of cometary origin. Thus the inter-comparison of the surface color variations and size distributions of Hildas/quasi-Hildas, JFCs and ACOs might give us some hints on the evolution of the surface material and structures of cometary nuclei, namely, from youth to old age in case the ACOs are representative of defunct (i.e., inactive) cometary nuclei. The aim of our observational project is therefore to establish a comprehensive data base of the sizes and surface colors of these populations of small bodies with a view to compare them to those of the Jupiter family comets (Lamy and Toth, 2009).

In this work we would bring attention to the serendipitous discovery of coma activity of an asteroid-in-cometary orbit, or ACO for short. This object, previously known as 2000 $YN_{30}$ (212P hereafter), was a target in our survey program of ACOs using the LOT one-meter telescope at Lulin Obervatory. It was first discovered by the NEAT (Near Earth Asteroid Tracking) group on Dec. 1, 2000 as an asteroid when it was at a solar distance of 1.86 AU just before perihelion. Its orbital parameters of a = 3.929 AU, e = 0.579 and i = 22.398$^o$ with $T_J$ = 2.635 make it a member of the ACOs. With an absolute magnitude H = 16.76 and a R-band geometric albedo $p_R$ of 0.096±0.032, its size can be estimated to be D = 1.7±0.3 km (Fernandez et al., 2005). That 212P has larger eccentricity and higher inclination than the quasi-Hildas or outliers listed in Toth (2006) might mean that it could actually be a weakly outgassing comet in transition to an inactive cometary nucleus. On the other hand, we note that two quasi-Hilda asteroids, 2004 $FM_{24}$ and 2002 $CF_{140}$ in Toth (2006), can be shown to evolve into cometary-like orbits within a million years by numerical calculations. Comet 212P might share the same dynamical origin as these two quasi-Hildas.

This paper is organized as follows. In Section 2 we will describe the observations and images obtained at Lulin. Section 3 will be dedicated to the discussion of the orbital evolution of 212P. Finally, the summary and discussion will be given in Section 4.

2. Observations

The LOT telescope used in this project was equipped with PI 1300B 1340x1300 pixels CCD camera. The image scale is 0.516" per pixel. The standard Asahi BVRI broadband filters were used in the photometric measurements. The observational log is given in Table 1. Figure 1 is the discovery image of 212P's outgassing activity taken on Jan $2^{nd}$, 2009 (Cheng et al., 2009). A faint dust tail with a length of about 20" or 14,000 km appeared in the anti-sunward direction. Note that 6.5 hours prior to our observation, A.R. Gibbs (IAUC 9010) found the presence of a diffuse coma surrounding this object that was not there in previous observational report in Dec, 2008 (MPEC-2008X77). After confirmation by IAU Circulars, this object was renamed 212P/2000 $YN_{30}$ (NEAT) to reflect its cometary nature.

Since the first detection of the dust tail feature, 212P was routinely monitored at Lulin until March of the same year. As shown in Figure 2, the images show consistently the formation of a faint dust tail. Morphologically speaking, the dust tail appears to have the same length between January 2 and January 7 and then becomes shorter and fainter afterwards. At the end of the Lulin observation in March, 2009, only a very diffuse structure could be recognized. An important question is thus whether the observed outgassing process could have existed more than just a few days.

This interpretation is also consistent with the finding that the Afρ (cm) value as determined according to A'Hearn et al.(1984) varies from 4.6 – 5.3 at the beginning of January, to 3.4 – 4.6, and finally to be about 2.8 in March as the comet moved away from the perihelion (see Table 2). The continuous presence of the dust tail between January and March, 2009, suggests uninterrupted emission process through this time interval according to the synchrone approach (Finson et al., 1968). However, we are unable to determine whether the dust emission was triggered by the recent excavation of some fresh active region by an impact event or not.

The next thing to check is about the colors of the dust coma. The aperture size of our photometric study is about 6~8 pixel (4200-5600 Km) depending on the seeing condition. The measured values of B-V (0.995±0.189), V-R (0.678±0.092) and V-I (1.008±0.106) at the optical center are consistent with those of the nuclei of JFCs (Lamy and Toth, 2009). Because of the faintness of the dust tail, it is not possible to estimate the color of the dust accurately. In any event, the B-R color at location away from the central nucleus by several arcsec in the tailward direction can be estimated to be about 1.1±0.1 (see Figure 3). This is consistent with the average value of about 1.1±0.3 for cometary dust as summarized in Kolokolova et al. (2004) and the B-R value of the dust tail of comet P/2010 TO20 LINEAR-Grauer which was found to be about 1.2±0.2 (Lacerda, 2012). Because the color remains nearly the same along the tail which brightness distribution has a much smaller slope as a function of the cometocentric distance than that of the gas coma, we believe that our measurements are not subject to strong contaminated by gas emission.

Figure 4 summarizes the orbital positions of 212P during the Lulin observations. The orbits of a number of main-belt comets are also shown for comparison. It can be seen that the coma activity occurred few days after perihelion and the dust coma remained detectable with LOT for the following two months. An enhanced level of the solar heating of the nucleus surface could be the triggering mechanism of the dust tail formation. Other effects might also play a role. For example, the absence of a dust coma when 212P was discovered as an asteroid might have two possible reasons. First, the appearance of a diffuse coma and dust tail in January 2009 could be associated with an outburst of a pocket of volatile material similar to that of comet 17P/Holmes (Lin et al., 2009; Reach et al., 2010) but at a smaller scale. Alternatively, the disappearance/non-detection of the dust features in earlier observations could be the result of seasonal effect connected to the obliquity of the rotating nucleus (Hsieh et al., 2010). It is therefore interesting to know whether 212P is basically a defunct comet with its nucleus surface covered by a dust mantle or that it is of the nature of a new comet that has reached the present perihelion distance only recently. We used numerical computation to trace the dynamical history of 212P.

3. Orbital Evolution

Table 3 shows the orbital parameters of 212P from Jet Propulsion Laboratory (JPL-SBD). We used the Mercury integrator of the Bulirsch-Stoer algorithm (Chambers, 1999) for the numerical orbit integrations. All planets from Mercury to Neptune are included in the computation. The step size employed was 0.05 day. The orbital evolution of 212P was traced backward and forward for 100,000 years in each direction.

Figure 5 shows an example of the pattern of orbital evolution according to the JPL ephemeris. In the case of the backward integration, 212P was captured into a "short-period" orbit with a < 20 AU at T=-46802 years ago as a result of a close encounter with Jupiter at a distance of 0.0063 AU. Its perihelion (q) stays at about 4 AU and aphelion (Q) at about 15 AU. Another close encounter with Jupiter at T=-18250 years ago at a distance of 0.021 AU transforms the orbit into a quasi-Hilda orbit with the semi-major axis being kept at about 4 AU. Our calculation shows that repeated planetary perturbations will lead to long-term oscillations in the eccentricity and inclination with periods varying between a few hundred years to about ten thousand years. At about t ~ -18250 years ago, the eccentricity reaches values as high as 0.8 allowing 212P attain a perihelion distance as close as 0.8 AU to the Sun. It is probably in this time interval that 212P begins its carrier as a short-period comet. This phase lasts about 250 years. Subsequently, the orbit of 212P will be transformed to one resembling more and more to that of a quasi-Hilda asteroid with q being raised to larger values between 1 and 3.5 AU. At the present time, the orbital evolution of 212P is in transition from a punctuated long-term 3:2 libration to a relatively smooth long-term libration in e and i for the next $10^5$ years. According to this sample calculation with the JPL orbital data, 212P will become a very stable quasi-Hilda object from now on.

The study of the orbital evolution of this test particle gives some idea on the possible origin and fate of an object like 212P. For example, it tells us how an object with perihelion distance originally outside the orbit of Jupiter could be transformed into a quasi-Hilda in long-term stable 3:2 resonance with Jupiter. However, the orbital elements given in the JPL ephemeris have numerical uncertainties. Furthermore, additional numerical effects could be introduced in the direct integration scheme. It is therefore of importance to study the statistical pattern of the orbital evolution of 212P by running a number of its clones. The Monte Carlo simulation of the orbital histories of 100 clones of 212P was done by following the description in Bernstein and Khushalani (2000).

The global pattern of the dynamical evolution of 212P -like objects could be formulated by computing the relative fractions of times in the lifetimes of individual test particles (clones) spent in different combinations of the orbital parameters. One method often used is the statistical distribution of the time intervals in the q

(perihelion) and Q (aphelion) coordinate system divided into many grids with Δq = 1 AU and ΔQ = 1 AU. The sum of the total time duration in each grid therefore constitutes the probability, namely, the distribution of residence time or "footprint" of the object during its orbital evolution.

Figure 6 compares the statistical distributions of the "footprints" of the 212P clones in the orbital space for the past and future one million years, respectively. The gravitational influences of Jupiter and Saturn are clearly seen. In the cases of the backward and forward integrations, we find several strips and clusters confined within the q-Q curve of Jupiter and then some others between those of Jupiter and Saturn, respectively. Similar features can be found between the q-Q curves of Saturn and Uranus. Those are objects being temporarily trapped between the orbits of these two outer planets. In the forward integration, there are also two horizontal patterns, one has q nearly being fixed at about 5 AU but Q moving from < 10 AU to > 1000 AU, and the other one has q being fixed at about 9 AU and Q ranging from 10 AU to beyond 1000 AU. These two tracks are produced by the gravitational scattering of the test particles by Jupiter and Saturn, respectively, to large aphelion distances. Similar "footprint" structures in the backward integration can be understood in the same manner. Because of their large gravitational scattering effect, both Jupiter and Saturn effectively form the barriers for further outward diffusion of the test particles to the orbital region of Uranus and Neptune. In the numerical simulations of injection of transneptuian objects into the inner solar system (see Tiscareno and Malhotra, 2003), we would find the presence of "footprints" between the orbit of Neptune and that of Uranus simply because of the fact that the source region is located in the vicinity of the trans-Neptunian region while Jupiter and Saturn form the barriers to inward diffusion.

Therefore, the evolutionary behavior as shown in Figure 6 is what would be expected of short-period comets which are injected from the trans-Neptunian region. However, in our backward integration, random planetary encounters tend to produce a cutoff at Jupiter's orbit thus curtailing the reverse paths returning to the trans-Neptunian region and the transitory zone occupied by Centaurs (Tiscareno and Malhotra, 2003). The orbital footprints from the forward integration show similar pattern as that of the backward integration.

These results also indicate that the exact evolutionary histories of quasi-Hilda asteroids are very sensitive to the starting values of the orbital elements. Even small differences could lead to very different outcomes in the orbital integrations. On the other hand, it is interesting to note that, in a statistical sense, some of the quasi-Hildas could have originated from the outer solar system via gravitational scattering by Jupiter. In this process, these objects including 212P would have moved in orbits with perihelia less than 1 AU for some short intervals. It means that there is a fair chance that 212P had experienced ice sublimation and gas outgassing from its surface material.

According to estimates of the average physical lifetime of short period comets (Fernandez, 1984; Hughes, 2003), the outgassing activity of those with perihelion q <1 AU should decay away with a physical lifetime of $10^3$-$10^4$ years because of the buildup of a dust mantle. For a short period comet orbit of a = 3 AU and e=0.8, the time spent inside 1AU is about 0.4 years per orbit with an orbital period of 5.2 years. This means that objects with a cumulative time of $\Delta t > 80 - 800$ years with q < 1 AU over its past orbital evolution would have a high probability of becoming inactive. Figure 7 shows the statistical distribution of the cumulative time of the 212P clones as a function of the heliocentric distance during their individual dynamical evolutions. Note that in our sample of "backward" test runs, a fraction (~30%) of the objects spend more than a total of $\Delta t \sim 100$ years inside 1 AU heliocentric distance. From this point of view, it is possible that 212P could have developed a partial dust mantle in its past history as a short-period comet.

4. Summary and Discussion

In the present work, we report the serendipitous discovery of coma activity of 212P which is a quasi-Hilda object in cometary-like orbit. This object with a diameter of D=1.7±0.3 km is of the size of a cometary nucleus. The results of our orbital integration show that it could have been captured into a short-period orbit from a Centaur-like orbit. Subsequent close encounters with Jupiter could have transformed its orbit into that of a quasi-Hilda characterized by the 3:2 mean motion resonance. The exact dynamical origin and its fate depend sensitively on the starting orbital elements. Our study therefore shows that ACOs/quasi-Hilda asteroids are of potential importance in tracing the transport process of volatile materials from the outer solar system to the inner solar system via planetary orbital migration and scattering. 212P will return to perihelion in 2016. Its brightness variation and possible reappearance of the dust coma activity will be closely monitored so that we can know for sure whether its observed dust tail structure last time is related to thermal sublimation or not. As for the larger issue of the Hildas and quasi-Hildas (and other classes of objects like the main-belt comets) as potential contributors to the terrestrial oceans, it is our intention to follow up with model calculations of their orbital evolution with and without the possible effects of planetary orbital migration. The definite answer would probably need to wait for in-situ (D/H) measurements (or sample returns) from these objects.

Acknowledgment.   We thank the reviewer for many useful and constructive comments which improve the scientific content of this work. We want to thank Yingtung Chen and Hsingwen Lin for useful discussions. This work was partially supported by NSC Grant: NSC 101-2119-M-008-007-MY3 and Ministry of Education under the Aim for Top University Program NCU, and Project 019/2010/A2 of Science and Technology Development Fund: MSAR. No. 0166.

Dello Russo, N., and Festou, M.C., 2008, LPICo, 1405.8216.

6. Figure Captions

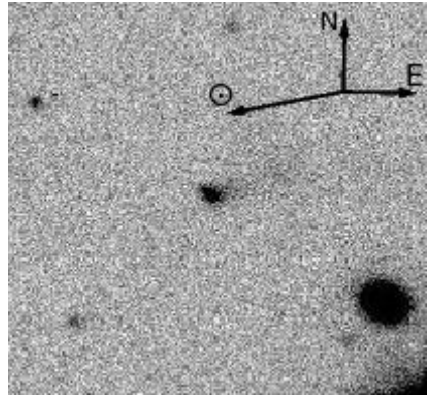

Figure 1. The discovery image of 212P taken at Lulin Observatory (B-band image with 300 second exposure).

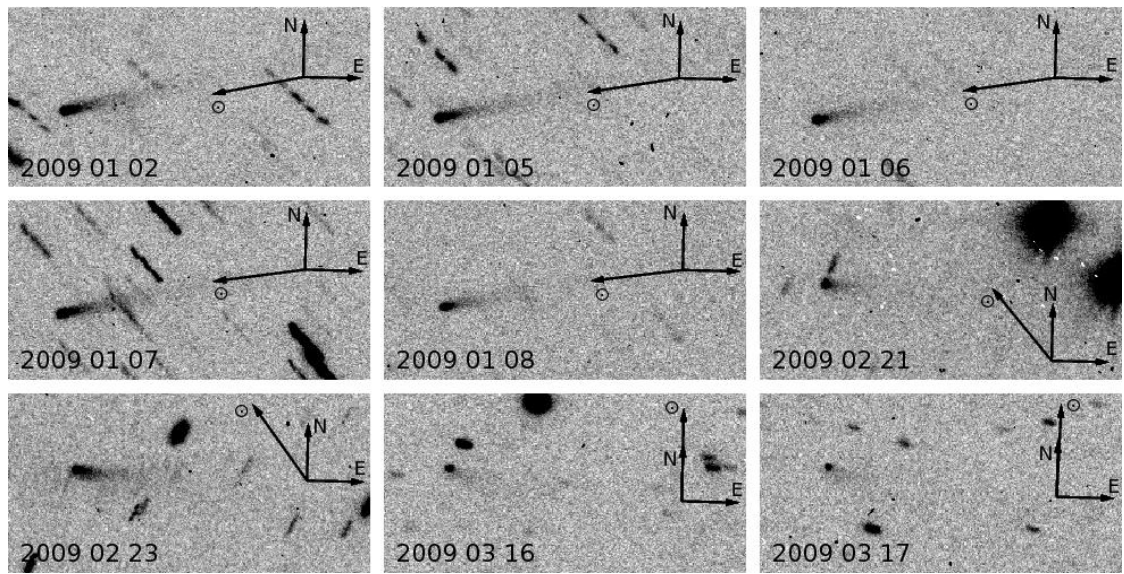

Figure 2. The follow-up images of 212P from January 2$^{nd}$ ,2009, until March 17, 2009. All the images were composites of three R-band images of 5 min. exposure. North is up and East is to the right. The direction of the sun is also indicated.

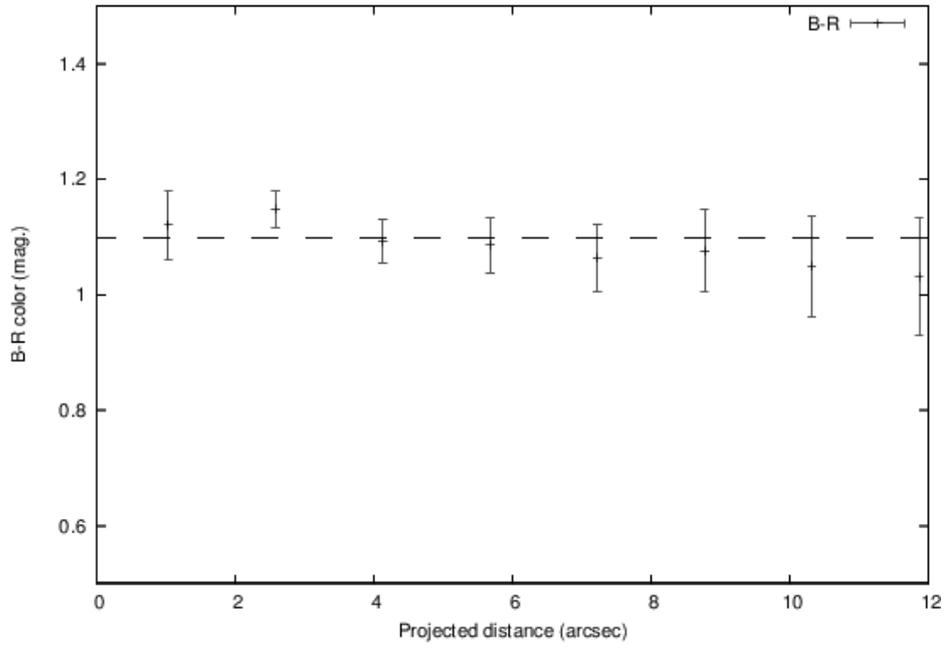

Figure 3. The B-R color of 212P along its dust tail on Jan. 2, 2009.

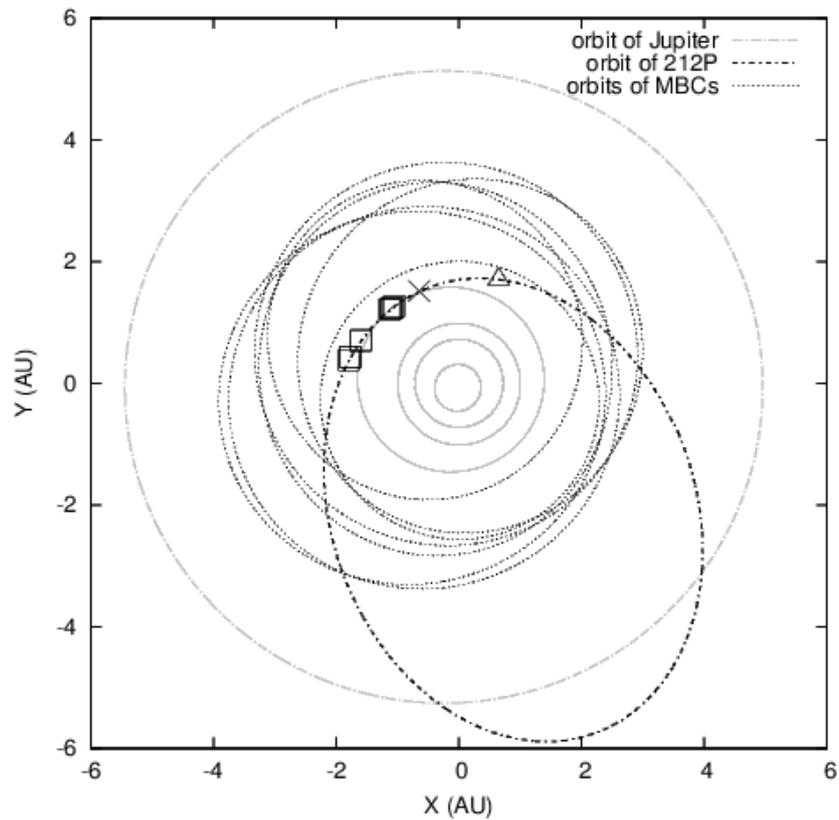

Figure 4. The orbit of 212P with some important positions identified: □ indicates the positions of our observations in 2009; Δ is for the orbital position of the first discovery in Dec, 2000. The orbits of several main-belt comets are also shown for comparison.

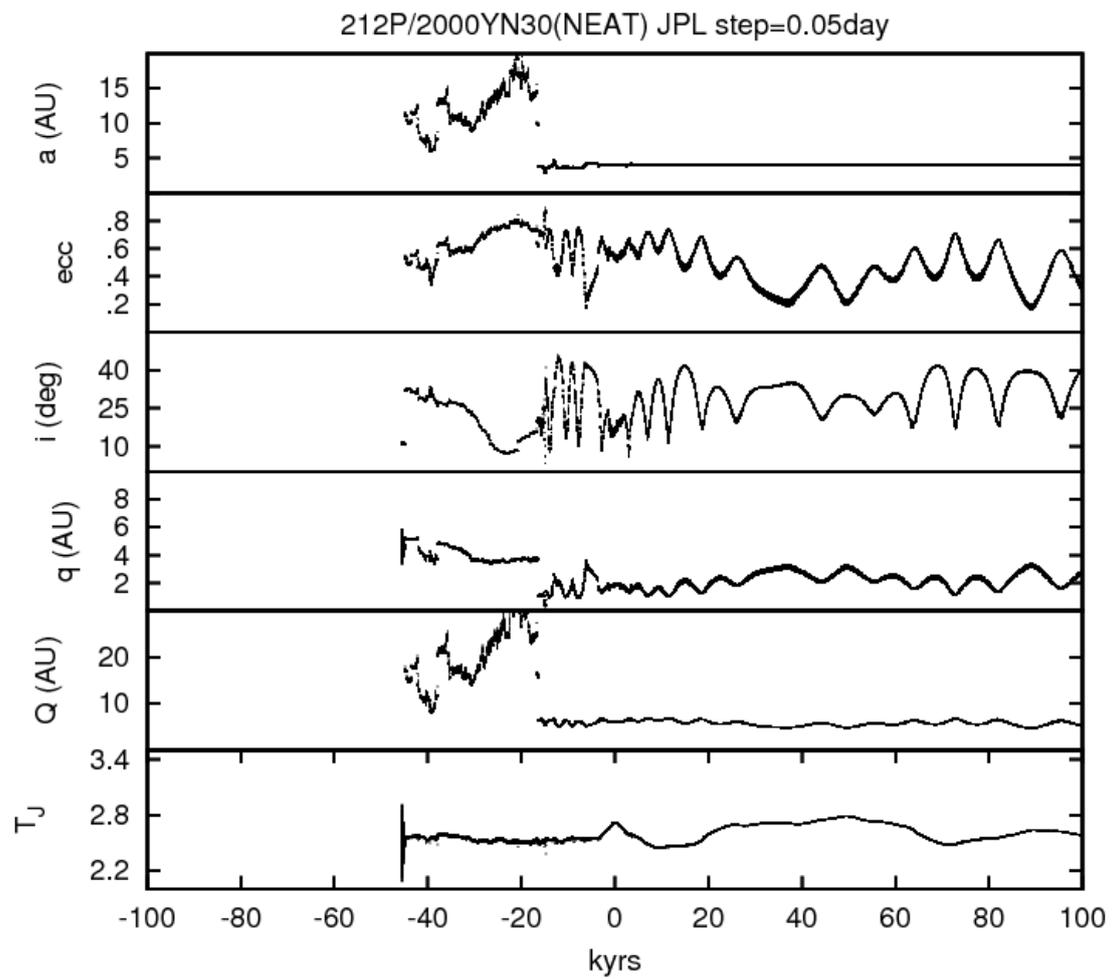

Figure 5. The time variations of the orbital parameters of 212P obtained by using initial values from JPL.

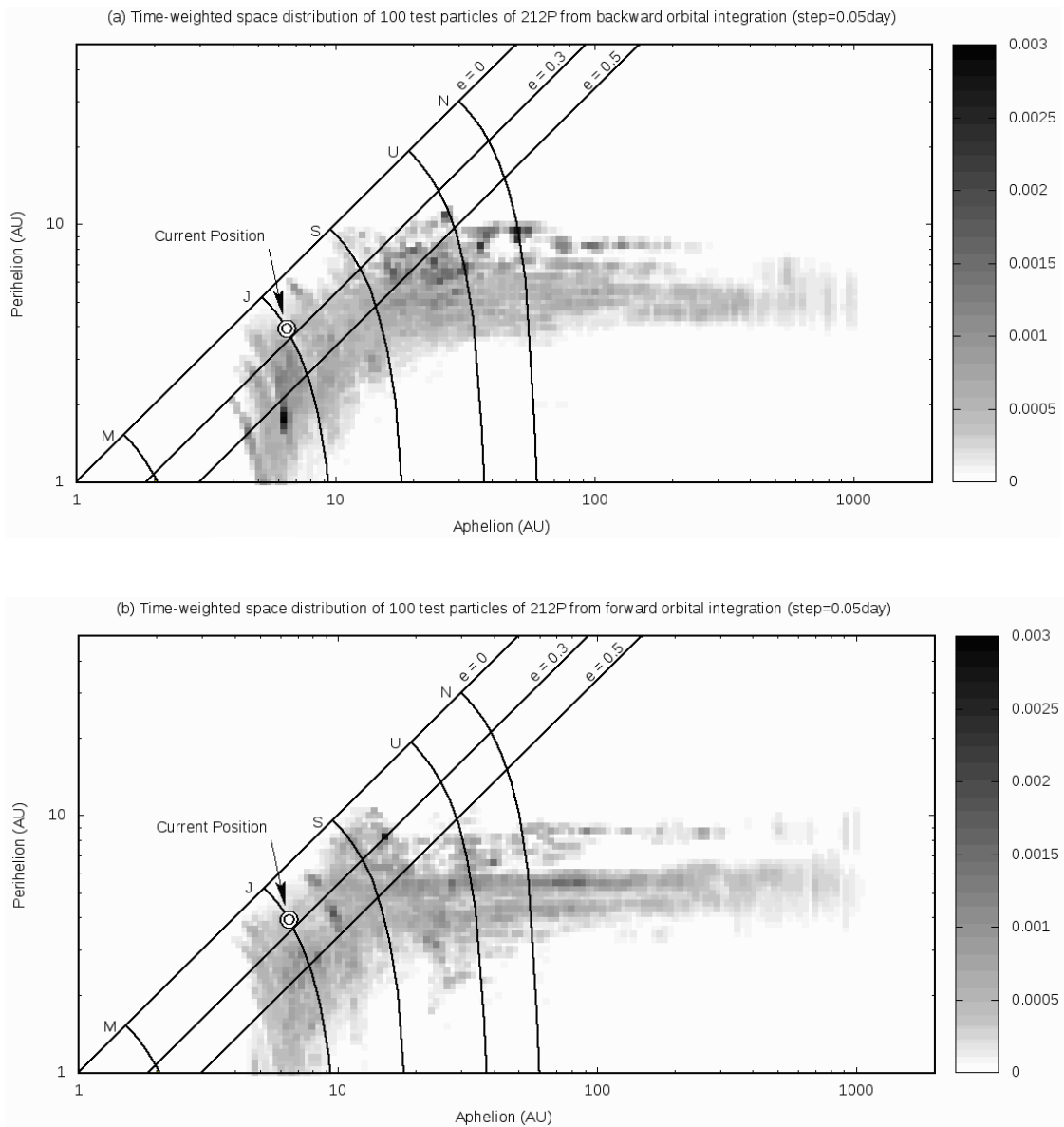

Figure 6. Probability distribution of the footprints of the 100 212P clones in the aphelion and perihelion distances. For our 1Myr integration, about 10% of the orbital duration would have the perihelion distances reaching below 2 AU. The current position is denoted by the black dot: (a) from backward orbital integration over the last $10^6$ years; (b) from forward orbital integration over the next $10^6$ years.

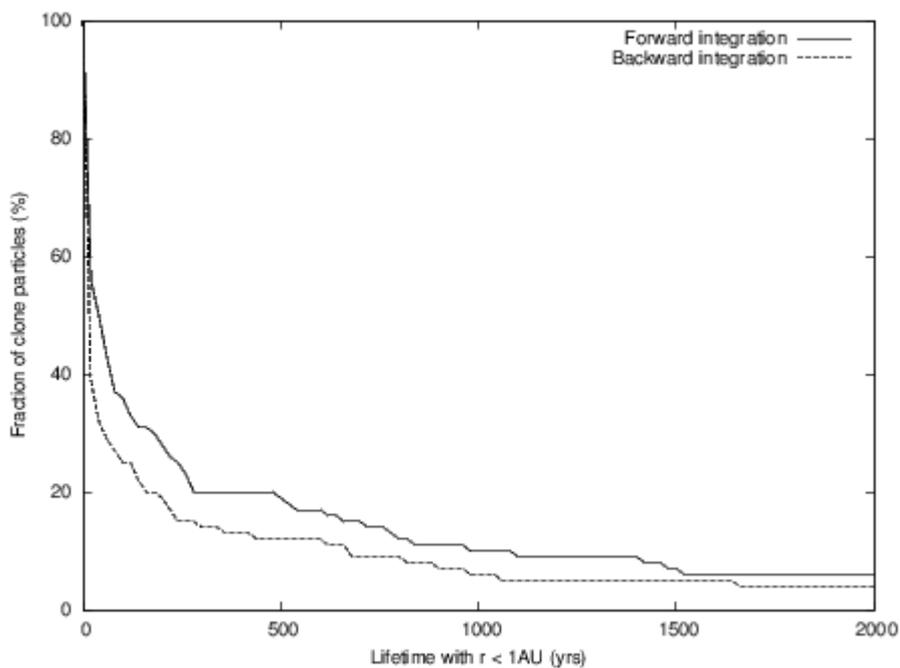

Figure 7. The statistical distributions of the cumulative time of the test bodies inside a certain heliocentric distance during their dynamical evolution in the backward (dashed lline) and forward (solid line) orbital integrations.

| Date | | $R_h$ (AU) | Δ(AU) | α(deg) | V | Remark |
|---|---|---|---|---|---|---|
| 2009 Jan | 2 | 1.683 | 0.959 | 30.4 | 19.594 (±0.064) | |
| | 5 | 1.688 | 0.948 | 29.6 | 19.548 (±0.070) | close to a BG stellar |
| | 6 | 1.690 | 0.944 | 29.4 | 20.027 (±0.094) | not photometric night |
| | 7 | 1.693 | 0.941 | 29.1 | 19.762 (±0.079) | |
| 2009 Mar | 17 | 1.950 | 1.107 | 20.7 | 20.700 (±0.130) | |
| | 18 | 1.955 | 1.115 | 20.8 | 20.512 (±0.129) | |
| | 19 | 1.960 | 1.123 | 20.9 | 20.504 (±0.132) | |

Table 1. The observation log of 212P by using the LOT telescope at Lulin Observatory.

| date | | rh (AU) | Δ (AU) | α (deg) | Afρ, 3.5" (cm) |
|---|---|---|---|---|---|
| 2009 Jan | 2 | 1.683 | 0.959 | 30.4 | 5.341 |
| | 5 | 1.688 | 0.948 | 29.6 | 4.712 |
| | 6 | 1.690 | 0.944 | 29.4 | 4.410 |
| | 7 | 1.693 | 0.941 | 29.1 | 4.927 |
| | 8 | 1.695 | 0.938 | 28.8 | 4.573 |
| 2009 Feb | 17 | 1.822 | 0.945 | 20.0 | 4.113 |
| | 21 | 1.839 | 0.961 | 19.7 | 3.385 |
| | 23 | 1.848 | 0.969 | 19.6 | 4.461 |
| 2009 Mar | 16 | 1.945 | 1.099 | 20.6 | 2.861 |
| | 17 | 1.950 | 1.107 | 20.7 | 2.844 |
| | 18 | 1.955 | 1.115 | 20.8 | 3.275 |
| | 19 | 1.960 | 1.123 | 20.9 | 2.707 |
| | 20 | 1.965 | 1.132 | 21.1 | 2.846 |

Table 2. The time variation of the Afρ values of 212P over the period of the Lulin observations in 2009.

| | JPL_SBD |
|---|---|
| Perihelion, q (AU) | 1.654234 |
| Eccentricity, e | 0.578748 |
| Inclination, i (deg.) | 22.3980 |
| Argument of pericentre (deg) | 15.0710 |
| Ascending node (deg) | 98.9294 |
| Epoch of pericentre, $T_p$ (JD) | 2454803.81008 |
| Orbital Epoch (JD) | 2454481.5 |

Table 3. Orbital parameters of 212P from NASA Jet Propulsion Laboratory.